\documentclass[aps,prl,twocolumn,showpacs]{revtex4-1}
\usepackage{graphicx,color}
\usepackage{amsmath}
\usepackage[colorlinks,citecolor=red]{hyperref}
\begin{document}

\let\up=\uparrow
\let\down=\downarrow
\newcommand\barI{\bar{I}}
\newcommand\avg[1]{\left\langle#1\right\rangle}

\title{Hanbury Brown and Twiss Correlations of Cooper Pairs in Helical
  Liquids}

\author{Mahn-Soo Choi}
\email{choims@korea.ac.kr}

\affiliation{Department of Physics, Korea University, Seoul 136-713, Korea}
\affiliation{School of Physics, Korea Institute for Advanced Study, Seoul 130-722, Korea}
\affiliation{Asia Pacific Center for Theoretical Physics, Pohang 790-784,
  Korea}

\date{\today}

\begin{abstract}
We propose a Hanbury Brown and Twiss (HBT) experiment of Cooper pairs on
the edge channels of quantum spin Hall insulators.
The helical edge channels provide a well defined beam of Cooper pairs and
perfect Andreev reflections from superconductors.  This allows our setup to be
identical in spirit to the original HBT experiment.
Interestingly, the cross correlation is \emph{always} negative and provides no
hint of the bosonic nature of Cooper pairs. This counter-intuitive result is
attributed to the perfect Andreev reflection and the true beam splitter in the
septup.
\end{abstract}
\maketitle

\definecolor{gray}{gray}{0.65}

\section{Introduction}

The Bose-Einstein condensation of constituent particles results in
superfluidity in bosonic systems.\cite{Pethick02a} Superconductivity in
electronic systems is attributed to the pairing of electrons into so called
``Cooper pairs''.\cite{Bardeen57b} It is thus plausible to expect Cooper pairs
to bear some bosonic nature, and to regard superconductivity as a
condensation.
Nevertheless, mathematically, Cooper pairs are not pure bosons because the
pair creation and annihilation operators do not obey strictly the boson
commutation relations,\cite{Bardeen57b}
and the bosonic nature of Cooper pairs
still remains controversial.  Therefore, it will be invaluable to examine the
issue directly in experiments.
Here we propose a Hanbury Brown and Twiss (HBT) experiment of Cooper pairs on
the edge channels of quantum spin Hall insulators,\cite{Bernevig06b,Konig07a}
a recently discovered new state of matter.
Surprisingly,\cite{endnote:1} the cross correlation is \emph{always} negative
and shows no hint of the bosonic nature of Cooper pairs once they are emitted
from the superconductor.
Previously, theoretical\cite{Anantram96a,Torres99a,Samuelsson02a,Bouchiat03a}
and experimental\cite{Hofstetter09a,Herrmann10a} works showed positive or
negative correlation depending on the system parameters.
Interestingly, the cross correlation has been predicted to be always negative
in a \emph{diffusive} multi-terminal superconductor-normal-metal
contacts.\cite{Nagaev01a}

A HBT effect is an \emph{intensity interference} between two partial beams.
It was originally introduced in order to overcome the technical difficulties
in measuring the size of stars with Michelson interferometers.  After the
pioneering experiment by HBT in 1956,\cite{HanburyBrown56a,HanburyBrown56b} it
was soon realized that the effect can determine quantum-statistical properties
of a stream of particles:\cite{Purcell56a,Glauber63c} The intensity
correlation is positive for bosons (obeying Bose-Einstein statistics) while
negative for fermions (obeying Fermi-Dirac statistics).\cite{Jeltes07a}

An ideal HBT experiment requires a well-defined beam of particles and a
tunable beam splitter (BS). For normal electrons, an electron beam is achieved
on the chiral edge states of an integer quantum Hall insulator (IQHI), for
which a quantum point contact (QPC) serves as a beam
splitter.\cite{Buttiker92a,Henny99a} It does not work for Cooper pairs because
the high magnetic field required for IQHI breaks them.

Quantum spin Hall insulator (QSHI) is a prototype example of topological
insulators.  Topological insulators are characterized by bulk excitation
energy gap and gapless edge modes, the latter being intimately connected to
the topological nature.\cite{Kane05a,Bernevig06b}
In QSHIs, the spin-orbit coupling gives the edge modes a helical feature, spin
up ($\up$) electrons moving to the right and spin down ($\down$) electrons
moving to the left, and the time-reversal invariance prevents back scattering
even in the presence of disorder.\cite{Kane05a,Bernevig06b} The helical edge
states are thus a Kramers pair and duplicate copies of a chiral edge mode.
Then it is clear that a QPC serves as a BS for helical edge
modes.\cite{Stroem09a,Hou09a}

When a superconductor are put on top of helical edge modes, the edge modes
become superconducting due to the proximity effect.\cite{Fu08a,Fu09a} The edge
modes beneath the superconductor will be referred to superconducting edge
modes or simply ``superconductors''.
A Cooper pair, itself being a pairing between time-reversal counterparts,
moves with no back scattering across the (point) interface between
superconductor and helical edge modes.  That is, the Andreev reflection is
\emph{always} perfect even in the presence of interface
barrier.\cite{endnote:2} This is in sharp contrast to usual
normal-superconductor hybrid structures, where in reality normal electron
reflection is unavoidable.  Therefore, the helical edge states provide a
well-defined and transparent channel for Cooper
pairs.\cite{endnote:3,*Alicea12a}
Two-dimensional QSHI has been observed recently in HgTe/CdTe quantum
wells\cite{Konig07a} and also expected in InAs/GaSb/AlSb type-II quantum
wells.\cite{Liu08a}

\begin{figure}
\centering
\includegraphics*[width=8cm]{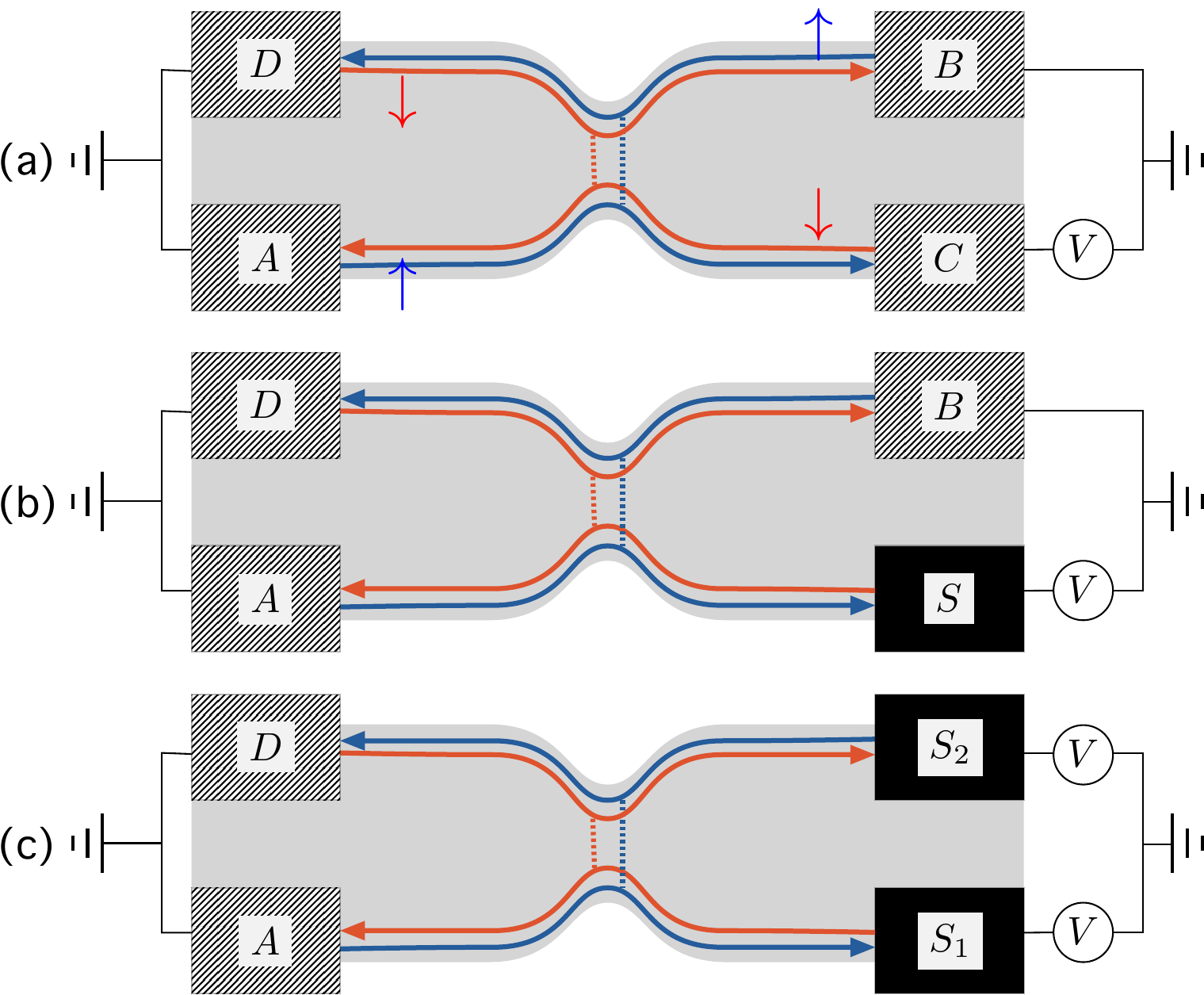}
\caption{The schematic setups for HBT experiment of Cooper pairs using edge
  states of QSHI.  The light gray represent the QSHI sample, the arrowed thick
  lines represent the helical edge states.  $\up$ ($\down$) electrons move
  counterclockwise (clockwise). The QPC formed by constriction with side gates
  serves as an electron beam splitter. A HBT experiment setup (a) for normal
  electrons and (b) for Cooper pairs. In (c), Cooper pairs are injected from
  two superconductors.}
\label{Paper::fig:1}
\end{figure}

Several HBT-\emph{type} experiments of Cooper pairs have been proposed
before.\cite{Anantram96a,Torres99a,Samuelsson02a,Bouchiat03a} In their cases,
however, a $Y$-junction is used instead of a true electron BS, and not only
Andreev but also normal-electron reflections are involved.\cite{Samuelsson02a}
It is shown below that these differences affect significantly the results.
Our setup is much closer in spirit to the original HBT experiment.
Another closely related work is Ref.~\onlinecite{Sato10a}: Here the Cooper pairs were injected to helical edge modes of QSHI through direct tunneling from spin-singlet superconductors. Unlike in our setup, the coupling between the helical edge modes and the superconductor in their setup is weak. As pointed out by the authors,\cite{Sato10a} most electrons are thus passing along the edges without Andreev scattering and rare electrons do undergo Andreev reflection but scatter to a different edge as holes. Such a crossed Andreev reflection leads to almost perfect positive correlation.\cite{Sato10a}

Recently, the splitting of Cooper pairs has been demonstrated
experimentally.\cite{Hofstetter09a,Herrmann10a} However, in these experiments
the scattering of Cooper pairs themselves have been suppressed with Coulomb
interaction.  While the experiments are only conductance measurement as they
are, the cross correlation is therefore expected manifestly positive in such
setups.\cite{Recher03a}

\begin{figure}
\centering
\includegraphics[width=8cm]{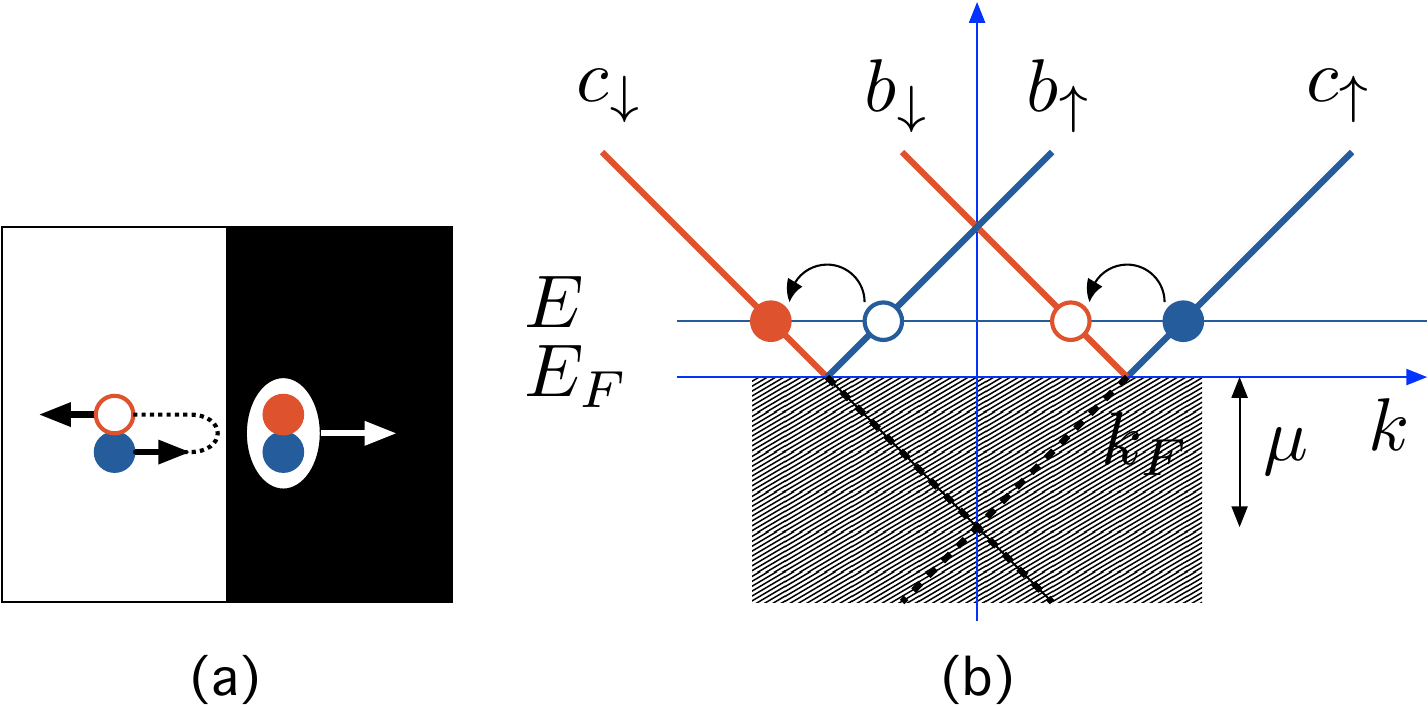}
\caption{(a) Andreev reflection at the interface between normal (left) and
  superconducting (right) helical edge modes, where $\up$ electron is
  reflected as a $\down$ hole. (b) The energy-momentum relation of particles
  ($c_\sigma$) and holes ($b_\sigma$) with spin $\sigma=\up,\down$.}
\label{Paper::fig:2}
\end{figure}

\section{Normal-Electron Case}

We consider a stripe of QSHI.  The helical edge states at its boundary
are described by a one-dimensional Dirac Hamiltonian
\begin{equation}
\label{Paper::eq:1}
H = \hbar{v}\sum_{\ell=1,2}\int{dx}\left[
  \psi_{\ell\down}^\dag(i\partial_x-\mu)\psi_{\ell\down} -
  \psi_{\ell\up}^\dag(i\partial_x+\mu)\psi_{\ell\up}
\right]
\end{equation}
where $\psi_\up$ ($\psi_\down$) is the field operator of $\up$ ($\down$)
electrons, $\mu$ is the chemical potential, and $v$ is the propagation
velocity.
The edge of a QSHI sample forms a closed loop and can never be terminated in
the middle.  In Eq.~(\ref{Paper::eq:1}) it has been assumed that the lower
($\ell=1$) and upper ($\ell=2$) segment of the whole edge are isolated from
the rest by putting contact reservoirs $A$, $B$, $C$, $D$, and other
additional contacts (not shown) as in Fig.~\ref{Paper::fig:1} (a).

Constricting the QSHI bar with side gates forms a QPC.
The $\up$ electron injected to the QPC from reservoir $A$, for example, either
moves on to reservoir $C$ with probability amplitude $t$ or tunnels to the
upper edge going out to reservoir $D$ with probability amplitude $r$.  The QPC
thus serves as a tunable BS for electrons.
The QPC is thus characterized completely by the scattering matrix $\Sigma$
\begin{equation}
\label{Paper::eq:2}
\begin{bmatrix}
c'_{1\down} \\ c'_{2\down} \\ c'_{1\up} \\ c'_{2\up} 
\end{bmatrix} = \Sigma
\begin{bmatrix}
c_{1\up} \\ c_{2\up} \\ c_{1\down} \\ c_{2\down} 
\end{bmatrix} =
\begin{bmatrix}
0 & 0 & t & r \\
0 & 0 & r & t \\
t & r & 0 & 0 \\
r & t & 0 & 0
\end{bmatrix}
\begin{bmatrix}
c_{1\up} \\ c_{2\up} \\ c_{1\down} \\ c_{2\down} 
\end{bmatrix}
\end{equation}
relating the amplitudes $c'_{\ell\sigma}$ ($\ell=1,2$ and $\sigma=\up,\down$)
at the output ports to $c_{\ell\sigma}$ at the input ports of QPC.
We have suppressed the energy (or momentum) dependence of
$c_{\ell\sigma}$ for notational simplicity.

We apply a bias voltage $V$ to reservoir $C$, keeping $A$, $B$, and $D$
electrically grounded. With this bias configuration, only $\down$ electrons
are injected from $C$ along the lower edge.  The quantum statistical
properties of the electron beam are characterized by the correlation functions
defined by
\begin{equation}
S_{\alpha\beta} = \int_{-\infty}^\infty{dt}
\avg{\Delta I_\alpha(t)\Delta I_\beta(0) + \Delta I_\beta(0)\Delta I_\alpha(t)}
\end{equation}
with $\Delta{I}_\alpha(t)=I_\alpha(t)-\barI_\alpha$, where $\barI_\alpha$ is
the average current \emph{into} reservoir $\alpha$.  Given the scattering
matrix in (\ref{Paper::eq:2}), the calculation of $\barI_\alpha$ and
$S_{\alpha\beta}$ is a simple application of the Landauer-B\"uttiker
formalism.\cite{Buttiker92a} We focus on the zero-temperature limit, $k_BT\ll
eV$.
In the present case, the average currents are given by
\begin{equation}
\label{Paper::eq:3}
\barI_A = \frac{e^2V}{2\pi\hbar}|t|^2 ,\quad
\barI_B = \frac{e^2V}{2\pi\hbar}|r|^2 .
\end{equation}
Clearly the injected current $\barI\equiv-\barI_C$ flows either to $A$ or $B$
($I_A+I_B=\barI$), but not to $D$ ($\barI_D = 0$).
This partitioning at the BS leads to the well-known
results\cite{Buttiker92a}
\begin{equation}
\label{Paper::eq:4}
S_{AA} = S_{BB} = -S_{AB} = 2e\barI|rt|^2
\end{equation}
for the current correlations, and its negative value is well understood by
their antibunching behavior.\cite{Buttiker92a} For later reference, we stress
that the contact $D$ is completely ``mute'', i.e., $I_D=0$ and
$S_{AD}=S_{BD}=S_{DD}=0$.

\section{HBT Correlations of Cooper Pairs}

Let us now turn to the setup, Fig.~\ref{Paper::fig:1} (b), of our main
concern.  We replace the normal contact $C$ by a superconducting contact $S$.
A superconducting edge state is achieved by putting an $s$-wave
superconductor on top of the edge.  Due to the proximity effect, the edge
states beneath the superconductor form a superconducting state with an induced
gap $\Delta=\Delta_0e^{-i\varphi}$.\cite{Fu09a} The Hamiltonian of the
superconducting edge modes is then given by [cf.~(\ref{Paper::eq:1})]
\begin{multline}
H_S = \hbar{v}\int{dx}\Big[
  \psi_{1\down}^\dag(i\partial_x-\mu)\psi_{1\down} -
  \psi_{1\up}^\dag(i\partial_x+\mu)\psi_{1\up} \\{}
  + \Delta\psi_{1\up}^\dag\psi_{1\down}^\dag
  + \Delta^*\psi_{1\down}\psi_{1\up}
\Big]
\end{multline}
We apply bias voltage $V$ to $S$ and keep $A,B$, and $D$ grounded, effectively
injecting Cooper pairs from $S$ along the lower edge channels.
The injection of Cooper pairs can be equivalently described by the Andreev
reflection process, where say an $\up$ \emph{electron} injected towards
superconductor is reflected by a $\down$ \emph{hole} away from superconductor,
see Fig.~\ref{Paper::fig:2}.  The Andreev reflection amplitude is given by
$a=e^{i\varphi-i\arccos(E/\Delta_0)}$ for incident electrons of energy $E$
($|E|<\Delta_0$). Note that due to the time-reversal invariance the Andreev
reflection probability remains unity, $|a|^2=1$, regardless of the
imperfections at the edge-superconductor interface.\cite{endnote:2} This is in
stark contrast to the usual normal-superconductor interface, where interface
imperfections suppress Andreev reflections.

Combining the Andreev reflections at the normal-superconductor interface and
the normal scattering (\ref{Paper::eq:2}) at the QPC, one obtains the
total scattering matrix for electrons and holes:
\begin{equation}
\label{Paper::eq:5}
\left[\begin{array}{c}
  c'_{1\down} \\ b'_{1\down} \\ 
  c'_{2\down} \\ b'_{2\down} \\ 
  c'_{2\up} \\ b'_{2\up}
  \end{array}\right] =
\left[\begin{array}{cccccc}
  0 & a^*|t|^2 & 0 & a^*r^*t & r & 0 \\
  a|t|^2 & 0 & art^* & 0 & 0 & r^* \\ 
  0 & a^*rt^* & 0 & a^*|r|^2 & t & 0 \\
  ar^*t & 0 & a|r|^2 & 0 & 0 & t^* \\ 
  r & 0 & t & 0 & 0 & 0 \\
  0 & r^* & 0 & t^* & 0 & 0
  \end{array}\right]
\left[\begin{array}{c}
  c_{1\up} \\ b_{1\up} \\ 
  c_{2\up} \\ b_{2\up} \\ 
  c_{2\down} \\ b_{2\down}
  \end{array}\right]
\end{equation}
where $b_{\ell\sigma}$ and $b'_{\ell\sigma}$ ($\ell=1,2$ and
$\sigma=\up,\down$) are amplitudes of the \emph{holes} on the input and output
ports.  Here $r^*\equiv[r(-E)]^*$, $t^*\equiv[t(-E)]^*$, and
$a^*\equiv[a(-E)]^*$ describe the scattering of \emph{holes}.  We ignore weak
energy dependence of $r$ and $t$.

The average currents $I_\alpha$ and the current correlations $S_{\alpha\beta}$
($\alpha,\beta=A,B,D$) are calculated using the Landauer-Buttiker formalism
extended to the normal-superconductor hybrid
structure.\cite{Beenakker97a,Blanter00a}
The currents
\begin{equation}
\barI_A = \frac{2e^2V}{2\pi\hbar}|t|^2 \,,\quad
\barI_B = \frac{2e^2V}{2\pi\hbar}|r|^2,
\end{equation}
are twice larger than (\ref{Paper::eq:3}), demonstrating perfect Andreev
reflections at (or injection of Cooper pairs from) $S$.  The correlation
functions are also given exactly in the same form as (\ref{Paper::eq:4}),
\begin{equation}
\label{Paper::eq:6}
S_{AA} = S_{BB} = -S_{AB} = 2e\barI|rt|^2 \,,
\end{equation}
except that the total current $\barI\equiv\barI_A+\barI_B$ is now twice
larger.  Surprisingly, the cross correlation is \emph{negative}.
(As in the normal case above, the contact $D$ is mute; $I_D=0$ and
$S_{AD}=S_{BD}=S_{DD}=0$).

\begin{figure}
\centering
\includegraphics*[width=6cm]{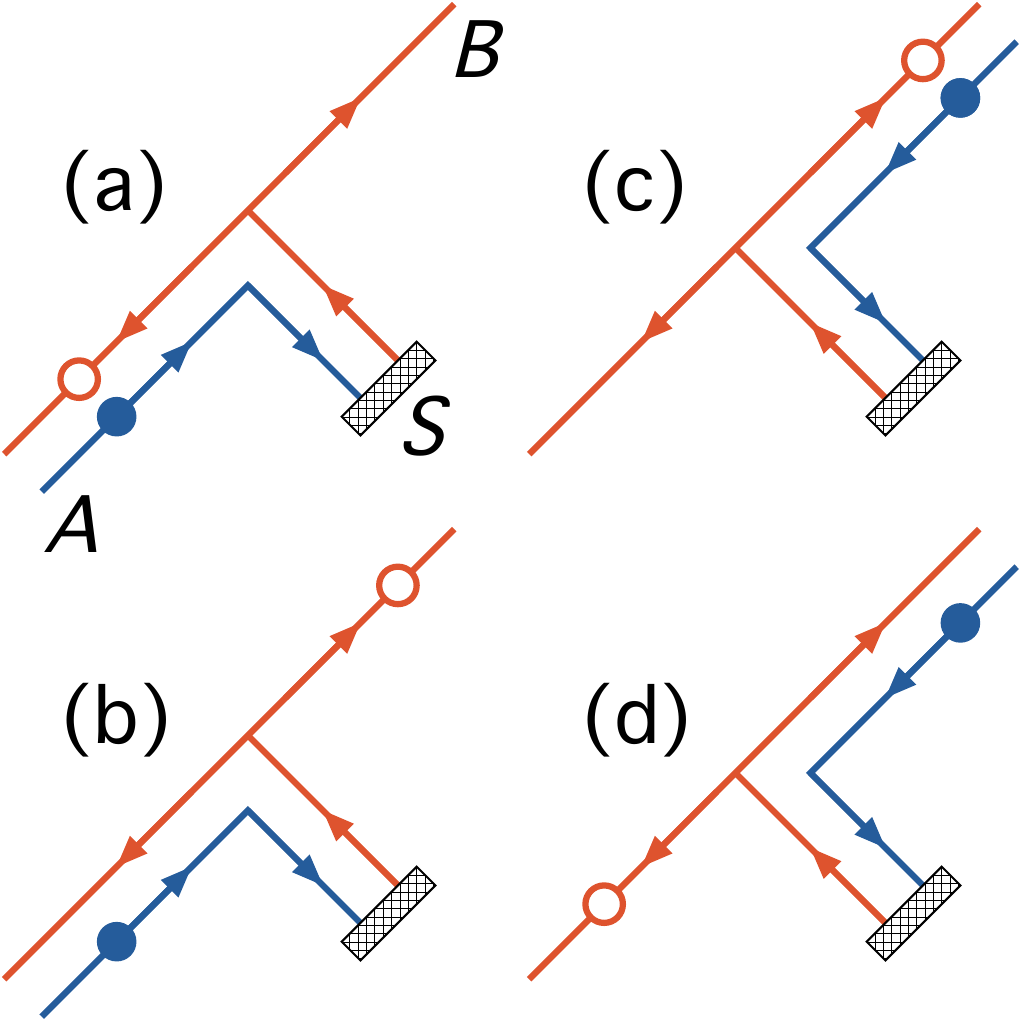}
\caption{Elementary scattering processes in a Hanbury Brown and Twiss
  experiment of Cooper pairs, described in electron-hole picture.  Blue (red)
  line represents the edge channels of $\up$ ($\down$) electrons (filled
  circle) or holes (empty circle).  In Cooper pair picture, panel (a)
  corresponds to Cooper pairs being injected from $S$, scattered at the
  quantum point contact (QPC), and going entirely to $A$.  Analogously, in
  (c), Cooper pairs entirely go to $B$.  In (c) and (d), constituent electrons
  in a Cooper pair split up at the QPC.}
\label{Paper::fig:3}
\end{figure}

Why is it surprising?  The description in~(\ref{Paper::eq:5}) in terms of
electrons and holes is equivalent to Cooper pairs injected from $S$ and
scattered at QPC.  Note that either (i) entire Cooper pairs go to $A$ or $B$
[Fig.~\ref{Paper::fig:3} (a) and (c)], or (ii) constituent electrons in each
pair split up into $A$ and $B$ [Fig.~\ref{Paper::fig:3} (b) and (d)].
Naively, one may expect a positive contribution from case (ii) with one
electron at each port $A$ and $B$ simultaneously.  Assuming (partial) bosonic
nature of Cooper pairs, one may also expect $S_{AD}>0$ in case (i).

The above naive expectation fails because it has ignored the two-particle
interference.\cite{Fano61b} For example, the two process ($A\up$-electron;
$B\up$-electron) $\to$ ($A\down$-hole; $B\down$-hole) [Fig.~\ref{Paper::fig:3}
(a) and (c)] and ($A\up$-electron; $B\up$-electron) $\to$ ($B\down$-hole;
$A\down$-hole) [Fig.~\ref{Paper::fig:3} (b) and (d)] are not distinguishable
and interfere with each other.  Due to the Fermi-Dirac statistics, the
amplitudes for these processes are opposite in sign, hence giving negative
cross correlation in~(\ref{Paper::eq:6}).

There is another simple way to understand the negative correlation.  For
example, $\up$ electron from $A$ undergoes either Fig.~\ref{Paper::fig:3} (a)
or (b).  Since the perfect Andreev reflection at the edge-superconductor
interface is noiseless, it is nothing but the partitioning of a single hole
and thus gives a negative contribution to $S_{AB}$.  Similarly, the
partitioning into Fig.~\ref{Paper::fig:3} (c) and (d) also give negative
contributions.

We thus conclude that Cooper pairs bear no bosonic characters at all, once
they get out of the superconductor.
It is interesting to note that in recent experiments with quantum
dots\cite{Hofstetter09a,Herrmann10a} the processes Fig.~\ref{Paper::fig:3} (a)
and (c) have been suppressed due to Coulomb interactions.  In this case, there
is no two-particle interference.

\section{Back-Scattering Effect}

The above conclusion appears contradictory to the previous theoretical
works,\cite{Anantram96a,Torres99a,Samuelsson02a,Bouchiat03a} where positive
correlation was predicted in certain range.
As mentioned earlier, in these works a $Y$ junction is used instead of a true
electron BS.  One crucial effect is the multiple reflections between the
junction and the superconductor.  Further, not only Andreev but also
normal-electron reflections are involved.  The latter effect cannot be
simulated in our system and is analyzed in the supplementary material.

To simulate the former effects, let us now replace both $C$ and $B$ by
superconductors, $S_1$ and $S_2$, with phases $\varphi_1$ and $\varphi_2$,
respectively.  We apply bias voltage $V$ on both $S_1$ and $S_2$ and keep $A$
and $D$ grounded.  The scattering of electrons and holes is governed by the
scattering matrix (ignoring corrections of order $E^2/\Delta_0^2\ll1$)
\begin{equation}
\renewcommand\arraystretch{1.5}
\begin{bmatrix}
c'_{1\down} \\ b'_{1\down} \\ c'_{2\up}\\ b'_{2\up}
\end{bmatrix} = \left[\begin{array}{cccc}
  0 & \frac{a_1^*|t|^2}{z^*} & \frac{w^*r}{z^*} & 0 \\
 \frac{a_1|t|^2}{z} & 0 & 0 & \frac{wr^*}{z} \\
 \frac{wr}{z} & 0 & 0 & \frac{a_2^*|t|^2}{z} \\
 0 & \frac{w^*r^*}{z^*} & \frac{a_2|t|^2}{z^*} & 0
 \end{array}\right]
\begin{bmatrix}
c_{1\up} \\ b_{1\up} \\ c_{2\down} \\ b_{2\down}
\end{bmatrix}
\end{equation}
where
\begin{math}
a_1=-e^{i\varphi_1+i2k_FL_1},
\end{math}
\begin{math}
a_2=-e^{i\varphi_2-i2k_FL_2},
\end{math}
$k_F=\mu/\hbar{v}$, and $L_1$ and $L_2$ are the distances from the QPC to the
superconducting contacts $S_1$ and $S_2$, respectively.
The \emph{multiple reflections} between QPC and superconductors are manifested
through the factors
\begin{math}
z = 1 - |r|^2e^{i\theta}
\end{math}
and 
\begin{math}
w = 1-e^{i\theta},
\end{math}
where 
\begin{math}
\theta=(\varphi_1-\varphi_2)+ 2k_F(L_1+L_2)
\end{math}
is the phase accumulation during one cycle of the multiple reflections.
This immediately leads to the resonance behavior of the average currents,
\begin{equation}
\barI_A = \barI_D = \frac{2e^2V}{2\pi\hbar}
\frac{|t|^4}{1+|r|^4-2|r|^2\cos\theta}
\end{equation}
and the correlations
\begin{equation}
S_{AA} = S_{DD} = S_{AD} = 2e\barI
\frac{|r|^2(1-\cos\theta)}{1+|r|^4-2|r|^2\cos\theta}
\end{equation}
For $\theta=2\pi{n}$ with $n$ an integer, the currents take maximum and the
noises vanish.

We note that the sign of the cross correlation $S_{AD}$ is now positive, in
contrast to (\ref{Paper::eq:6}), which is always negative.  This difference
is ascribed to the multiple reflections between QPC and superconductors.  As
noted by Baym,\cite{Baym74a} the intensity correlation $\avg{I(t+\tau)I(t)}$
is equivalent to the relative probability to observe two particles at two
points separated by a distance $v\tau$, where $v$ is the flight velocity.  If
the HBT experiment is done with a ``true BS'' as in
Fig.~\ref{Paper::fig:1} (b),
then the intensity correlation is entirely due to the spatial distribution and
the quantum statistical property of the particles in the beam itself.  To the
contrary, if the BS is replaced by a
$Y$-junction\cite{Anantram96a,Torres99a,Samuelsson02a,Bouchiat03a} and
multiple reflections occur between the source and the junction, then the cross
correlation is not directly related to the spatial distribution in the beam
alone but affected significantly by the successive interactions with the
source.
Therefore, the positive cross correlations in
Refs.~\cite{Anantram96a,Torres99a,Samuelsson02a,Bouchiat03a} do not represent
entirely the quantum statistical properties of Cooper pairs \emph{after}
emitted from superconductor (away by distance larger than the superconducting
coherence length).




\begin{acknowledgments}
This work was supported by the NRF grant (2011-0012494) and the BK21 program.
The author expresses thanks to R. Aguado, A. Levy Yeyati, and R. Zambrini for
useful discussions, and special thanks to M. B\"uttiker for helful comments
and clarifying a few issues.
\end{acknowledgments}

\appendix
\section{Comparison of the Chaotic Quantum Dot and the Helical Liquid}

Previous theoretical works\cite{Samuelsson02a,Samuelsson02b} considered a chaotic quantum dot coupled to
two normal leads 2 and 1 (corresponding to $A$ and $B$ in our case) and one
superconducting reservoir.  They found
both positive and negative cross correlations in a wide range of parameter
values, and provide clear interpretations of the relevant microscopic
processes.  Therefore, it will be useful for a deeper understanding of the
physics behind the cross correlation to compare more closely the processes in
their and our setup.

The crucial differences are (i) the multiple reflections between the
dot-superconductor interface and the $Y$-junction (dot-normal-metal
interface), and (ii) the \emph{normal-electron reflections} (other than
Andreev reflections).  With the former effect discussed in the main text of
the paper, here we focus on the latter effect.

Here we follow the scattering-matrix approach in \cite{Samuelsson02a}, which
provides a clear picture of the two-particle interference \citep{Fano61b} and
hence direct comparison of their system to ours.  A semiclassical analysis is
given in \cite{Samuelsson02b}, which is more useful for interpretation in
terms of partitioning noise.

We first discuss the case with the transparency $\Gamma_S$ of the
dot-superconductor is perfect ($\Gamma_S=1$).  The normal and superconducting
contacts supports $N$ and $M$ channels, respectively.

\begin{figure}[h]
\centering
\includegraphics[width=6cm]{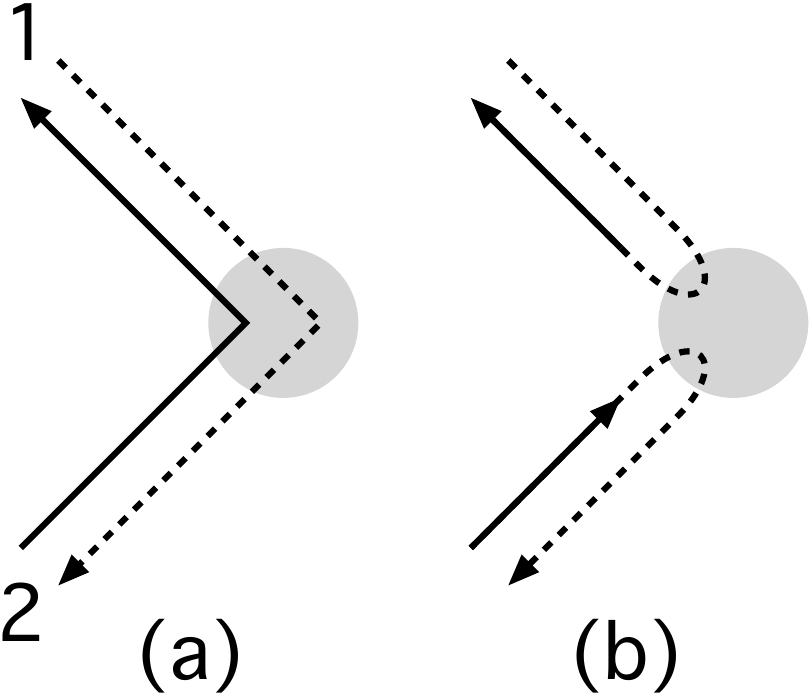}
\caption{Equation~(\ref{Supplementary::eq:1}) corresponds to the interference of
  the two processes (a) and (b).  The interference is positive since it only
  involves the exchange of electron and hole, i.e., different species of
  particles.}
\label{Supplementary::fig:1}
\end{figure}

For small $2N/M\ll 1$,
the dominant process is the (local) Andreev reflections at the normal
contact-dot interface, due to the gap in the dot spectrum induced by the
superconductor.  For finite (but small) $2N/M$, there are finite probabilities
of normal reflections.
The cross Andreev reflections (CAR) from one normal reservoir to the other is
still negligible; i.e., the terms such as
$(S^{eh}_{12})$ can be ignored.  In this limit, the cross correlation is given
by
\begin{equation}
\label{Supplementary::eq:1}
\frac{P_{12}}{4e^3V/2\pi\hbar}
\approx \frac{P_{12}^{eh}+P_{12}^{he}}{4e^3V/2\pi\hbar}
\approx
2\operatorname{Tr}\left[(S^{ee}_{12})^\dag(S^{eh}_{11})(S^{hh}_{21})^\dag(S^{he}_{22})\right]
\end{equation}
This term arises from the two-particle interference of the two processes (a)
and (b) in Fig.~\ref{Supplementary::fig:1}: The two processes cannot be
distinguished and therefore the amplitudes (not probability) should be summed.
The interference gives \emph{positive} cross correlation \citep{Fano61b}
since it involves the exchange of electron and hole, namely, different species
of particles.  Therefore, the normal scattering processes (either electron or
hole) in Fig.~\ref{Supplementary::fig:1} is crucial to the positive cross
correlation in this regime.  In our case, the process of
Fig.~\ref{Supplementary::fig:1} is prohibited because of the helical (chiral)
nature of the edge states.

\begin{figure}[h]
\centering
\includegraphics[width=6cm]{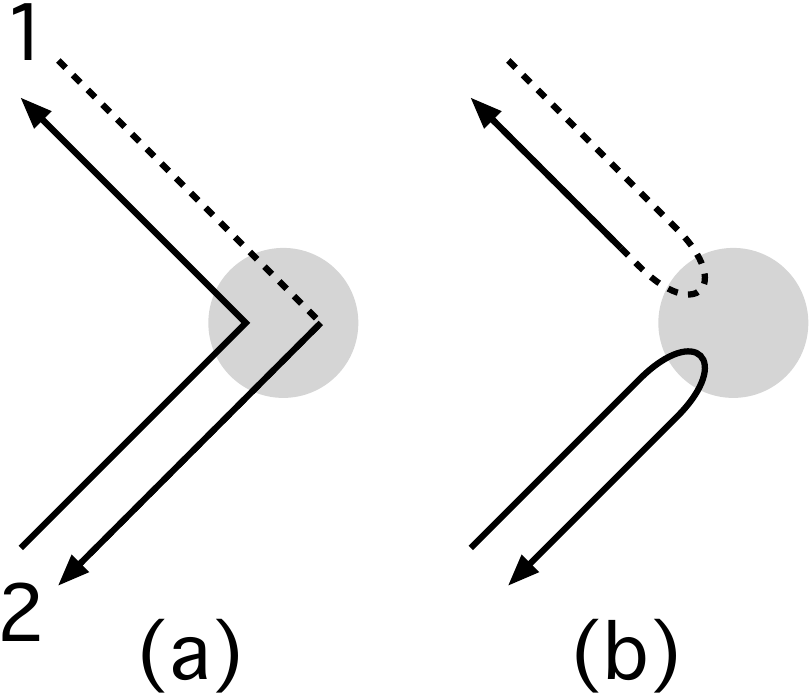}
\caption{The third term in Equation~(\ref{Supplementary::eq:2}),
  $\operatorname{Tr}(S^{ee}_{12})^\dag(S^{eh}_{11})
  (S^{eh}_{21})^\dag(S^{ee}_{22})$,
  corresponds to the interference of the two processes (a) and (b).  The
  interference is negative since it involves electrons (i.e., the electrons
  after the AR).}
\label{Supplementary::fig:2}
\end{figure}

In the limit of $2N/M\gg 1$, \cite{Samuelsson02a} obtained negative cross
correlation.  In this limit, there are finite probabilities of the CAR
processes.  According to \cite{Samuelsson02a}, the main contributions come
from $P_{12}^{ee}$ and $P_{12}^{hh}$, where $P_{12}^{ee}$ is expressed
explicitly as
\begin{widetext}
\begin{multline}
\label{Supplementary::eq:2}
-\frac{P_{12}^{ee}}{4e^3V/2\pi\hbar}
= \operatorname{Tr}\left[(S^{ee}_{11})^\dag(S^{eh}_{11})(S^{eh}_{21})^\dag(S^{ee}_{21})
  + (S^{ee}_{11})^\dag(S^{eh}_{12})(S^{eh}_{22})^\dag(S^{ee}_{21})\right] \\{}
+ \operatorname{Tr}
\left[(S^{ee}_{12})^\dag(S^{eh}_{11})(S^{eh}_{21})^\dag(S^{ee}_{22})
  + (S^{ee}_{12})^\dag(S^{eh}_{12})(S^{eh}_{22})^\dag(S^{ee}_{22})\right]
\end{multline}
\end{widetext}
For example, the third term
\begin{math}
\operatorname{Tr}
\left[(S^{ee}_{12})^\dag(S^{eh}_{11})(S^{eh}_{21})^\dag(S^{ee}_{22})\right]
\end{math}
corresponds to the two-particle interference of the two processes
Fig.~\ref{Supplementary::fig:2} (a) and (b).  The interference is
\emph{negative} in this case, because it involves the exchange of two
electrons (after the Andreev reflection).  Again, both processes
Fig.~\ref{Supplementary::fig:2} (a) and (b) include \emph{normal electron
  reflections}, which are forbidden in our case.  Other terms also involve
similar normal electron reflections.  The above analysis shows that the system
considered in \cite{Samuelsson02a,Samuelsson02b} is in a clear distinction
from ours, which involves only Andreev reflections (scattering of Cooper
pairs).

Let us now turn to the case with $\Gamma_S<1$.  It turns out that for
$\Gamma_S<1/2$, the cross correlation becomes positive in the limit
$2N/M\Gamma_S\gg 1$.  This is ascribed to the additional fluctuations at the
dot-superconductor interface due to imperfect Andreev reflections, which is
forbidden in our system due to the time-reversal symmetry and makes another
difference.

It will be useful to simulate this latter effect in our system by applying
a magnetic field, which effectively introduces a mass gap into the helical
liquid \citep{Fu09a}.

In conclusion, the close comparisons above reveals where the discrepancy
between their and our results arise.  While in our setup the cross correlation
involves only the scattering of Cooper pairs (or Andreev reflections), in
\cite{Samuelsson02a} it involves normal electron reflections as well as
Andreev reflections.

\bibliographystyle{physrev}
\bibliography{Paper}

\begin{thebibliography}{10}
\bibitem{endnote:3}
In helical edge modes, the spin-$\up$ (spin-$\down$) electron moves only to the
  right (left). Cooper-pair injection is still poissible: When a Cooper pair is
  inject, it has a finite center-of-momentum, say, $q$ so that the momenta of
  two constituting electrons are $q+k$ and $q-k$. One should also recall the
  $p$-wave nature of the pairing in helical edge modes originating from strong
  spin-orbit coupling.
\end{thebibliography}

\end{document}